\documentclass[12pt]{article}

\begin{document}
\begin{center}
{\bf Higher Derivative Scalar Field Theory in the First Order Formalism}\\
\vspace{5mm}
 S. I. Kruglov \\
\vspace{5mm}
\textit{University of Toronto at Scarborough,\\ Physical and Environmental Sciences Department, \\
1265 Military Trail, Toronto, Ontario, Canada M1C 1A4}
\end{center}

\begin{abstract}

The scalar field theory with higher derivatives is considered in
the first order formalism. The field equation of the forth order
describes scalar particles possessing two mass states. The first
order relativistic wave equation in the 10-dimensional matrix form
is derived. We find the relativistically invariant bilinear form
and corresponding Lagrangian. The canonical energy-momentum tensor
and density of the electromagnetic current are obtained. Dynamical
and non-dynamical components of the wave function are separated
and the quantum-mechanical Hamiltonian is found. Projection
operators extracting solutions of field equations for definite
energy and different mass states of particles are obtained. The
canonical quantization of scalar fields with two mass states is
performed, and propagators are found in the formalism considered.

\end{abstract}

\section{Introduction}
It is known that the gravity theory based on the Einstein-Hilbert
action is non-renormalizable in four dimensions \cite{Hooft}. By
including curvature squared terms in the action \cite{Stelle}, the
theory becomes renormalizable but the higher derivative (HD)
theory. There are also HD fields in generalized electrodynamics
\cite{Podolski}. HD theories allow us to improve renormalization
of the theory and to avoid ultraviolet divergences
\cite{Thirring}. However, it was discovered soon that there are
some difficulties with negative norm (ghosts), and unitarity in HD
theories \cite{Pais}, \cite{Heisenberg}. Much attention,
therefore, was made to study the simplest version of the HD theory
of scalar fields ( see \cite{Urries}, \cite{Polonyi},
\cite{Hawking}, \cite{Rivelles}, \cite{Smilga} and references
therein). It should be mentioned that scalar fields play very
important role in the theory of inflation of the universe. So,
inflation scalar fields can be candidates of dark matter of the
universe. HD scalar fields are also present in SUSY field theories
in extra dimensions (see, for example, \cite{Kazakov}). Although
HD field theories lead to ghosts, it was shown that problems with
negative probabilities and S-matrix unitarity can be solved
\cite{Hawking}. The purpose of this paper is to formulate the
forth order equation for scalar fields in the form of the first
order relativistic wave equation, to obtain the Lagrangian,
conserved currents, the quantum-mechanical Hamiltonian, and to
perform the quantization.

The paper is organized as follows. In Sec. 2, we derive the first
order relativistic wave equation for scalar fields in the
10-dimensional matrix form, the relativistically invariant
bilinear form and the Lagrangian formulation. In Sec. 3, the
canonical energy-momentum tensor and the electromagnetic current
density are obtained. Dynamical and non-dynamical components of
the wave function are separated and the quantum-mechanical
Hamiltonian is found in Sec. 4. In Sec. 5 projection operators
extracting solutions of field equations for definite energy and
different mass states of particles are obtained. The canonical
quantization of scalar fields with two mass states is performed,
and the propagators of scalar fields were found in the formalism
of the first order in Sec. 6. We discuss results obtained in Sec.
7. In Appendix, some useful products of matrices are given. The
system of units $\hbar =c=1$ is chosen, Greek and Latin letters
run $1,2,3,4$ and $1,2,3$, correspondingly.

\section{Scalar Field Equation of the Forth Order}

Consider the forth order field equation describing scalar
particles possessing two mass states \cite{Urries},
\cite{Hawking}:
\begin{equation}
\left( \partial^2
 -m_1^2 \right)\left(\partial^2 -m_2^2 \right)\varphi(x)=0 ,
\label{1}
\end{equation}
where $ \partial^2 \equiv\partial_\nu^2$, $\partial_\nu
=\partial/\partial x_\nu =(\partial/\partial x_m,\partial/\partial
(it))$. It is obvious that Eq. (1) has two solutions corresponding
to mass $m_1$ and $m_2$.

Let us introduce the 10-dimensional wave function
\begin{equation}
\phi (x)=\left\{ \phi _A(x)\right\} =\left(
\begin{array}{c}
\varphi (x)\\
\widetilde{\varphi}(x)\\
\varphi _\mu (x)\\
\widetilde{\varphi}_\mu (x)
\end{array}
\right) \hspace{0.5in}(A=0 , \widetilde{0} ,\mu ,\widetilde{\mu})
,\label{2}
\end{equation}
where $\phi_0 (x)=\varphi (x)$, $\phi_{\widetilde{0}}
(x)=\widetilde{\varphi} (x)$, $\phi_\mu (x)=\varphi_\mu (x)$,
$\phi_{\widetilde{\mu}} (x)=\widetilde{\varphi}_\mu (x)$,
\[
\widetilde{\varphi} (x)=\frac{1}{m_1 m_2}\partial^2 \varphi
(x),~~~~ \varphi _\mu (x)=\frac{1}{m_1+m_2}\partial_\mu \varphi(x)
,
\]
\vspace{-7mm}
\begin{equation} \label{3}
\end{equation}
\vspace{-7mm}
\[
 \widetilde{\varphi}_\mu(x)=\frac{1}{m_1+m_2}\partial_\mu
\widetilde{\varphi}(x) .
\]
The function $\phi (x)$ represents the direct sum of two scalars
$\varphi (x)$, $\widetilde{\varphi}(x)$, and two four-vectors
$\varphi _\mu (x)$, $\widetilde{\varphi}_\mu (x)$.

Introducing the elements of the entire matrix algebra $\varepsilon
^{A,B}$ (see, for example \cite{Kruglov}) with matrix elements and
products
\begin{equation}
\left( \varepsilon ^{M,N}\right) _{AB}=\delta _{MA}\delta _{NB},
\hspace{0.5in}\varepsilon ^{M,A}\varepsilon ^{B,N}=\delta
_{AB}\varepsilon ^{M,N}, \label{4}
\end{equation}
where $A,B,M,N=0,\widetilde{0},\mu,\widetilde{\mu}$, Eq. (1) can
be cast in the form of the first order equation
\[
\partial _\mu \left[\varepsilon ^{0,\widetilde{\mu}}- \varepsilon ^{\widetilde{0},\mu}
- \sigma\varepsilon^{0,\mu}-\frac{m}{m_1+
m_2}\left(\varepsilon^{\mu,0}+\varepsilon^{\widetilde{\mu},
\widetilde{0}}\right)\right]_{AB}\phi _B(x)
\]
\vspace{-7mm}
\begin{equation} \label{5}
\end{equation}
\vspace{-7mm}
\[
+ m\left[\varepsilon ^{0,0}+\varepsilon
^{\widetilde{0},\widetilde{0}}+ \varepsilon ^{\mu ,\mu }+
\varepsilon ^{\widetilde{\mu} ,\widetilde{\mu} }\right] _{AB}\phi
_B(x)=0 , \label{5}
\]
where
\[
m=\frac{m_1 m_2}{m_1+m_2},~~~~\sigma=\frac{m_1^2+m_2^2}{m_1 m_2} ,
\]
and there is a summation over all repeated indices. We define
10-dimensional matrices as follows:
\begin{equation}
\rho _\mu =\varepsilon ^{0,\widetilde{\mu}}- \varepsilon
^{\widetilde{0},\mu} - \sigma\varepsilon^{0,\mu}-\frac{m}{m_1+
m_2}\left(\varepsilon^{\mu,0}+\varepsilon^{\widetilde{\mu},
\widetilde{0}}\right) ,
 \label{6}
\end{equation}
\begin{equation}
I_{10}=\varepsilon ^{0,0}+\varepsilon
^{\widetilde{0},\widetilde{0}}+ \varepsilon ^{\mu ,\mu }+
\varepsilon ^{\widetilde{\mu} ,\widetilde{\mu} } ,\label{7}
\end{equation}
where $I_{10}$ is a unit 10-dimensional matrix. Then Eq. (5)
becomes the relativistic wave equation of the first order:
\begin{equation}
\left( \rho _\mu \partial _\mu +m\right) \phi (x)=0 . \label{8}
\end{equation}
So, we reformulated the higher derivative equation for scalar
fields (1) in the form of the first order Eq. (8). Now, one can
apply general methods to investigate the first order matrix
equation \cite{Gel'fand}. The spectrum of the particle mass of Eq.
(8) is given by $m/\lambda_i$, where $\lambda_i$ are the
eigenvalues of the matrix $\rho_4$. It is not difficult to verify
that the matrix
\[
\rho _4 =\varepsilon ^{0,\widetilde{4}}- \varepsilon
^{\widetilde{0},4} - \sigma\varepsilon^{0,4}-\frac{m}{m_1+
m_2}\left(\varepsilon^{4,0}+\varepsilon^{\widetilde{4},
\widetilde{0}}\right)
\]
satisfies the matrix equation
\begin{equation}
\rho _4^4-\frac{m_1^2+m_2^2}{(m_1+m_2)^2}\rho_4^2 +
\frac{m_1^2m_2^2}{(m_1+m_2)^4}\Lambda = 0 , \label{9}
\end{equation}
where
\begin{equation}
\Lambda =\varepsilon ^{0,0}+\varepsilon
^{\widetilde{0},\widetilde{0}}+ \varepsilon ^{4 ,4 }+ \varepsilon
^{\widetilde{4} ,\widetilde{4} } \label{10}
\end{equation}
is the projection operator extracting four-dimensional subspace,
$\Lambda^2=\Lambda$. As the eigenvalue of the matrix $\Lambda$ is
unit, we obtain from Eq. (9) eigenvalues of the matrix $\rho_4$ as
follows:
\begin{equation}
\lambda_1=\pm\frac{m_2}{m_1+m_2},~~\lambda_2=\pm\frac{m_1}{m_1+m_2}
. \label{11}
\end{equation}
So, positive masses of scalar particles described by the first
order Eq. (8) are $m/\lambda_1=m_1$, $m/\lambda_2=m_2$.

The Lorentz group generators in the representation space are
\cite{Kruglov}
\begin{equation}
J_{\mu\nu}= \varepsilon^{\mu,\nu}-\varepsilon^{\nu,\mu} +
\varepsilon^{\widetilde{\mu},\widetilde{\nu}}-\varepsilon^{\widetilde{\nu},\widetilde{\mu}}
,
 \label{12}
\end{equation}
and obey the commutation relations
\begin{equation}
\left[ J_{\mu \nu },J_{\alpha \beta}\right] =\delta _{\nu \alpha
}J_{\mu \beta}+\delta _{\mu \beta }J_{\nu \alpha}-\delta _{\nu
\beta }J_{\mu \alpha}-\delta _{\mu \alpha }J_{\nu \beta} .
\label{13}
\end{equation}
The relativistic form-invariance of Eq. (8) follows from the
relationship
\begin{equation}
\left[ \rho _\lambda ,J_{\mu \nu }\right] =\delta _{\lambda \mu
}\rho _\nu -\delta _{\lambda \nu }\rho _\mu  . \label{14}
\end{equation}
One may verify with the help of Eq. (4), (6), (12) that Eq. (14)
is valid.

The Hermitianizing matrix $\eta$ has to obey the relations
\cite{Gel'fand}
\begin{equation}
\eta \rho _m=-\rho _m^{+}\eta^{+} ,\hspace{0.5in}\eta \rho _4=\rho
_4^{+}\eta^{+} \hspace{0.5in}(m=1,2,3) .  \label{15}
\end{equation}
We obtain
\[
\eta=\varepsilon^{0,0}-\varepsilon^{\widetilde{0},\widetilde{0}}-
\frac{\sigma(m_1+m_2)}{m}\left(\varepsilon^{m,m}-\varepsilon^{4,4}
\right)
\]
\vspace{-7mm}
\begin{equation} \label{16}
\end{equation}
\vspace{-7mm}
\[
+\frac{(m_1+m_2)}{m}\left(\varepsilon^{m,\widetilde{m}} +
\varepsilon^{\widetilde{m},m} - \varepsilon^{4,\widetilde{4}} -
\varepsilon^{\widetilde{4},4}\right) .
\]
The matrix $\eta $ is the Hermitian matrix, $\eta^{+}=\eta$.
Introducing the matrix $\overline{\phi }(x)=\phi ^{+}(x)\eta$
($\phi ^{+}(x)$ is the Hermitian-conjugate wave function), one
obtains from Eq. (8) the ``conjugate" equation
\begin{equation}
\overline{\phi }(x)\left( \rho _\mu \overleftarrow{\partial} _\mu
-m\right) =0 . \label{17}
\end{equation}
The Lagrangian is given by the standard equation
\begin{equation}
{\cal L}=-\frac{1}{2}\left[\overline{\phi }(x)\left(\rho _\mu
\partial _\mu +m\right) \phi (x)-\overline{\phi }(x)\left(\rho _\mu
\overleftarrow{\partial} _\mu -m\right) \phi (x)\right]
,\label{18}
\end{equation}
so that the relativistically invariant bilinear form is $
\overline{\phi }(x)\phi (x)=\phi ^{+}(x)\eta \phi(x)$.

With the aid of Eq. (2),(3),(16), the Lagrangian (18) becomes
\[
{\cal L}=-\frac{1}{m_1+m_2}\biggl\{\frac{1}{2m_1
m_2}\left[\varphi^{*}
\partial^4\varphi +\left(\partial
^4\varphi^{*}\right)\varphi\right]
\]
\vspace{-7mm}
\begin{equation} \label{19}
\end{equation}
\vspace{-7mm}
\[
-\frac{\sigma}{2}\left[\varphi^{*}
\partial^2\varphi+\left(\partial^2\varphi^{*}
\right)\varphi\right]+m_1 m_2 \varphi^{*} \varphi \biggr\} .
\]
Up to total derivatives, which do not influence on an equation of
motion, the Lagrangian (19) takes the compact form
\[
{\cal L}=-\frac{1}{m_1 m_2(m_1+m_2)}\biggl[
\left(\partial_\mu\partial_\nu\varphi^{*}\right)
\left(\partial_\mu\partial_\nu\varphi\right)
\]
\vspace{-7mm}
\begin{equation} \label{20}
\end{equation}
\vspace{-7mm}
\[
+ \left(m_1^2+m_2^2\right)\left(\partial_\mu\varphi^{*}\right)
\left(\partial_\mu\varphi\right) +m_1^2 m_2^2 \varphi^{*} \varphi
\biggr] .
\]
One may verify that the Euler-Lagrange equation \cite{Barut}
\begin{equation}
\frac{{\partial\cal L}}{\partial\varphi^{*}}-\partial_\mu
\frac{\partial{\cal L}}{\partial(\partial_\mu\varphi^{*})} +
\partial_\mu
\partial_\nu\frac{\partial{\cal L}}{\partial(\partial_\mu\partial_\nu\varphi^{*})}
=0, \label{21}
\end{equation}
for the higher derivative Lagrangian (20), leads to the field
equation (1).

\section{The Energy-Momentum Tensor and Electromagnetic Current}

With the help of the general expression \cite {Ahieser}
\begin{equation}
T_{\mu\nu}=\frac{\partial\mathcal{L}}{\partial\left(\partial_\mu
\phi (x)\right)}\partial_\nu \phi (x)+\partial_\nu \overline{\phi
}(x) \frac{\partial\mathcal{L}}{\partial\left(\partial_\mu
\overline{\phi} (x)\right)}-\delta_{\mu\nu} \mathcal{L}
,\label{22}
\end{equation}
we obtain from the Lagrangian (18) the canonical energy-momentum
tensor
\begin{equation}
 T_{\mu\nu}=\frac{1}{2}\left(\partial_\nu \overline{\phi}
(x)\right)\rho_\mu \phi (x)-\frac{1}{2} \overline{\phi}
(x)\rho_\mu \partial_\nu\phi (x) . \label{23}
\end{equation}
It was taken into account that $ \mathcal{L}=0$ for functions
$\phi (x)$, $\overline{\phi} (x)$ satisfying Eq. (8), (17). Using
Eq. (2)-(4), (6), one finds from Eq. (23) the expression as
follows:
\[
T_{\mu\nu}=\frac{1}{2(m_1+m_2)}\biggl\{ \left(m_1^2+m_2^2\right)
\left[\varphi^{*}\partial_\mu\partial_\nu\varphi-\left(\partial_\mu\varphi^{*}\right)
\partial_\nu\varphi\right]-\varphi^{*}\partial_\mu\partial_\nu\partial^2 \varphi
\]
\vspace{-7mm}
\begin{equation} \label{24}
\end{equation}
\vspace{-7mm}
\[
-\left(\partial^2\varphi^{*}\right)\partial_\mu\partial_\nu
\varphi +\left(\partial_\mu\varphi^{*}\right)
\partial_\nu\partial^2 \varphi+
\left(\partial_\nu\varphi^{*}\right)
\partial_\mu\partial^2 \varphi \biggr\}+ c.c. ,
\]
where c.c. means the complex conjugate expression. The energy
density and the momentum density are given by ${\cal E}=T_{44}$,
$P_m=iT_{m4}$.

The electric current density is \cite{Ahieser}
\begin{equation}
j_\mu (x)=i\left( \overline{\phi }(x)
\frac{\partial\mathcal{L}}{\partial\left(\partial_\mu
\overline{\phi} (x)\right)}-
\frac{\partial\mathcal{L}}{\partial\left(\partial_\mu \phi
(x)\right)}\phi (x)\right) . \label{25}
\end{equation}
Replacing Eq. (18) into Eq. (25), one obtains the electric current
density
\begin{equation}
j_\mu (x)=i\overline{\phi }(x)\rho_\mu \phi(x) , \label{26}
\end{equation}
so that $\partial_\mu j_\mu (x)=0$. With the help of Eq. (2)-(4),
(6), we find
\[
j_\mu=\frac{i}{m_1m_2(m_1+m_2)}\biggl\{ \left(m_1^2+m_2^2\right)
\left[\left(\partial_\mu\varphi^{*}\right) \varphi-
\varphi^{*}\partial_\mu\varphi\right]
\]
\vspace{-7mm}
\begin{equation} \label{27}
\end{equation}
\vspace{-7mm}
\[
+ \varphi^{*}\partial_\mu\partial^2
\varphi-\left(\partial_\mu\partial^2\varphi^{*}\right) \varphi+
\left(\partial^2\varphi^{*}\right)
\partial_\mu \varphi -\left(\partial_\mu\varphi^{*}\right)
\partial^2\varphi\biggr\} .
\]
For the real scalar fields, $\varphi=\varphi{*}$, the electric
current vanishes.

\section{Quantum-Mechanical Hamiltonian}

Introducing an interaction of scalar particles with
electromagnetic fields by the substitution $\partial _\mu
\rightarrow D_\mu =\partial _\mu -ieA_\mu $, where $A_\mu $ is the
four-vector potential of electromagnetic fields (the minimal
electromagnetic interaction), Eq. (8) may be represented as
follows:
\begin{equation}
i\rho _4\partial _t\phi (x)=\biggl [\rho_a D_a+m + eA_0\rho_4
\biggr ]\phi(x).  \label{28}
\end{equation}
Let us introduce two auxiliary functions
\begin{equation}
\Xi (x)=\Lambda \phi(x) ,~~ \Omega (x)=\Pi\phi(x) ,  \label{29}
\end{equation}
where the projection operator $\Lambda$ is given by Eq. (10), and
the projection operator $\Pi$ is
\begin{equation}
 \Pi=1-\Lambda=\varepsilon ^{m,m}+\varepsilon
^{\widetilde{m},\widetilde{m}} ,  \label{30}
\end{equation}
so that $\Xi (x)+\Omega (x)=\phi(x)$. Acting on Eq. (28) by the
operator
\[
\frac{(m_1+m_2)^2}{m_1^2 m_2^2}\rho_4\left[ m_1^2+m_2^2
-\left(m_1+m_2\right)^2 \rho_4^2 \right],
\]
and taking into consideration Eq. (9), we obtain
\[
i\partial _t \Xi (x)=eA_0 \Xi (x)
\]
\vspace{-7mm}
\begin{equation} \label{31}
\end{equation}
\vspace{-7mm}
\[
+ \frac{(m_1+m_2)^2}{m_1^2 m_2^2}\rho_4\left[ m_1^2+m_2^2
-\left(m_1+m_2\right)^2 \rho_4^2 \right]\left(\rho_a D_a +
m\right)\phi (x) .
\]
Multiplying Eq. (28) by the operator $\Pi$, one finds
\begin{equation}
\Pi\rho_a D_a\Xi(x) + m\Omega(x)=0 . \label{32}
\end{equation}
We took into account here that $\Pi\rho_4=0$, $\Pi\rho_m \Pi=0$.
It follows from Eq. (31),(32) that the dynamical components of
wave function are given by the $\Xi (x)$, and non-dynamical
components by the $\Omega(x)$. The $\Xi (x)$ possesses four
components, and the auxiliary function $\Omega(x)$ has six
components. To separate the dynamical and non-dynamical
components, we express the auxiliary function $\Omega(x)$ from Eq.
(32), and replace it into Eq. (31). As a result, one obtains the
equation in the Hamiltonian form:
\begin{equation}
i\partial _t\Xi (x)=\widehat{\mathcal{H}}\Xi (x) , \label{33}
\end{equation}
\[
\widehat{\mathcal{H}}=eA_0 + \frac{(m_1+m_2)^2}{m_1^2 m_2^2}\left[
m_1^2+m_2^2 -\left(m_1+m_2\right)^2 \rho_4^2 \right]\rho_4
\]
\vspace{-7mm}
\begin{equation} \label{34}
\end{equation}
\vspace{-7mm}
\[
\times\left(\rho_a D_a + m\right)\left(1-\frac{1}{m}\Pi\rho_m D_m
\right).
\]
Two components of the function $\Xi (x)$ correspond to the state
with the mass $m_1$ (with positive and negative energy) and the
other two - to the state with the mass $m_2$.  With the help of
products of matrices, given in Appendix, the Hamiltonian (34)
becomes:
\[
\widehat{\mathcal{H}}=eA_0 -\sigma m \varepsilon
^{\widetilde{4},\widetilde{0}}-\left(m_1+m_2\right)\left(
\varepsilon ^{0,4}+\varepsilon
^{\widetilde{0},\widetilde{4}}\right)+m\left(\varepsilon
^{\widetilde{4},0}-\varepsilon ^{4,\widetilde{0}}\right)
\]
\vspace{-7mm}
\begin{equation} \label{35}
\end{equation}
\vspace{-7mm}
\[
+\frac{1}{m_1+m_2}\left(\varepsilon
^{\widetilde{4},\widetilde{0}}+\varepsilon ^{4,0}\right)D_m^2 .
\]
We have omitted the linear term in derivatives $L\equiv\left(
\varepsilon ^{4,m}+\varepsilon
^{\widetilde{4},\widetilde{m}}\right)D_m$ in the Hamiltonian,
because $L\Xi(x)=0$. Taking into consideration Eq.
(2),(4),(10),(35), we can rewrite Eq. (33) in the component form
\[
i\partial _t\varphi = eA_0 \varphi -\left( m_1+m_2\right)\varphi_4
,~~~~i\partial _t\widetilde{\varphi}= eA_0
\widetilde{\varphi}-\left( m_1+m_2\right)\widetilde{\varphi}_4 ,
\]
\begin{equation}
i\partial _t\varphi_4 = eA_0 \varphi_4-m\widetilde{\varphi}
+\frac{1}{m_1+m_2}D_m^2\varphi , \label{36}
\end{equation}
\[
i\partial _t\widetilde{\varphi}_4 = eA_0
\widetilde{\varphi}_4-\sigma m\widetilde{\varphi}+m\varphi
+\frac{1}{m_1+m_2}D_m^2\widetilde{\varphi} .
\]
The system of equations (36) may be obtained from Eq. (1), (3),
after the exclusion of components
\[
\varphi_m=\frac{1}{m_1+m_2}\partial_m\varphi ,~~~~
\widetilde{\varphi}_m=\frac{1}{m_1+m_2}\partial_m
\widetilde{\varphi} .
\]
According to Eq. (33) only components with time derivatives enter
Eq. (36), and describe the evolution of fields in time.

\section{Mass Projection Operators}

Consider solutions of Eq. (8) with definite energy and momentum
for two mass states, $\tau=1,2$,  in the form of plane waves:
\begin{equation}
\phi_\tau^{(\pm)}(x)=\sqrt{\frac{m_\tau^2}{p_0 V m}}v_\tau(\pm
p)\exp(\pm ipx) , \label{37}
\end{equation}
where $V$ is the normalization volume, $p^2=-m_\tau^2$ (no
summation in index $\tau$). We imply that four momentum
$p=(\textbf{p},ip_0)$ possesses the additional quantum number
$\tau=1,2$ corresponding two masses. The 10-dimensional function
$v_\tau(\pm p)$ obeys the equation
\begin{equation}
\left(i\hat{p}\pm  m \right)v_\tau(\pm p)=0 , \label{38}
\end{equation}
where $\hat{p}=\rho_\mu p_\mu$. It is convenient to use the
normalization conditions
\begin{equation}
\int_V \overline{\phi}^{(\pm)}_{\tau}(x)\rho_4 \phi^{(\pm)}_{\tau
}(x)d^3 x=\pm1 ,~~~~\int_V \overline{\phi}^{(\pm)}_{\tau}(x)\rho_4
\phi^{(\mp)}_{\tau }(x)d^3 x=0 , \label{39}
\end{equation}
where $\overline{\phi}^{(\pm)}_{\tau}(x)=\left(\phi^{(\pm)}_{\tau
}(x)\right)^+ \eta$. Normalization conditions (39) lead to relations
for functions $v_\tau(\pm p)$:
\begin{equation}
\overline{v}_\tau(\pm p)\rho_\mu v_\tau(\pm
p)=\mp\frac{imp_\mu}{m_\tau^2} ,~~~~\overline{v}_\tau(\pm p)
v_\tau(\pm p)=1 . \label{40}
\end{equation}
It is not difficult to verify, with the help of Eq. (4), (6) (see
Appendix), that the minimal equation for the matrix $\hat{p}$ is
\begin{equation}
\hat{p}^5
-\frac{\left(m_1^2+m_2^2\right)p^2}{\left(m_1+m_2\right)^2}
\hat{p}^3 + \frac{m^2 p^4 }{\left(m_1+m_2\right)^2} \hat{p}=0 .
\label{41}
\end{equation}
Now we obtain the projection matrices corresponding to definite
energy, momentum, and quantum number $\tau$:
\begin{equation}
\Pi_\tau (\pm p)=\frac{\left(m_1+m_2\right)^4 i \hat{p}\left(
i\hat{p}\mp m\right)}{2m_1^2m_2^2\left(m_\tau^4-m_1^2m_2^2\right)}
\left[ \widehat{p}^2+
\frac{m_\tau^4}{\left(m_1+m_2\right)^2}\right] . \label{42}
\end{equation}
The projection operators (42) obey equations as follows:
\begin{equation}
\left(ip\pm  m \right)\Pi_\tau(\pm)=0 , \label{43}
\end{equation}
\begin{equation}
\Pi_\tau(\pm p)^2=\Pi_\tau(\pm p) ,~~~~\Pi_\tau(+p)\Pi_\tau(-p)=0,
~~~~\mbox{tr}\Pi_\tau(\pm p)=1 .\label{44}
\end{equation}
Projection matrices (42) can be represented (see \cite{Fedorov})
as matrix-dyads
\begin{equation}
\Pi_\tau (\pm p)=v_\tau(\pm p)\cdot \overline{v}_\tau(\pm p) ,
\label{45}
\end{equation}
so that matrix elements of matrix-dyads are
\[
\left(v_\tau(\pm p)\cdot \overline{v}_\tau(\pm
p)\right)_{MN}=\left(v_\tau(\pm p) \right)_M
\left(\overline{v}_\tau(\pm p)\right)_N.
\]
Projection operators (42) extract solutions of Eq. (38) for
definite energy and different mass states of particles. Eq. (42),
(45) allow us to calculate matrix elements of different processes
of interactions of scalar particles in the covariant form.

\section{ Field Quantization }

One can obtain the momenta from Eq. (18):
\begin{equation}
\pi (x)=\frac{\partial\mathcal{L}}{\partial(\partial_0\phi
(x)}=\frac{i}{2}\overline{\phi}(x)\rho_4
 ,\label{46}
\end{equation}
\begin{equation}
\overline{\pi}(x)=\frac{\partial\mathcal{L}}{\partial(\partial_0
\overline{\phi}(x))} =-\frac{i}{2}\rho_4\phi(x) , \label{47}
\end{equation}
where the fields $\phi(x)$, $\overline{\phi} (x)$ are independent
``coordinates". The Hamiltonian density is given by the equation
\[
 {\cal H}=\pi (x)\partial_0\phi (x)+\left(\partial_0\overline{\phi}
 (x)\right)\overline{\pi} (x) -{\cal L}
\]
\vspace{-7mm}
\begin{equation} \label{48}
\end{equation}
\vspace{-7mm}
\[
=\frac{i}{2}\overline{\phi}(x)\rho_4\partial_0\phi (x)
 -\frac{i}{2}\left(\partial_0\overline{\phi}(x)\right)\rho_4\phi
 (x),
\]
It follows from Eq. (23) that the value ${\cal H}=T_{44}$ is the
energy density.

In the quantized theory the field operators can be written as
follows
\[
\phi_\tau(x)=\sum_{p}\left[a_{\tau,p}\phi^{(+)}_{\tau}(x) +
b^+_{\tau,p}\phi^{(-)}_{\tau}(x)\right] ,
\]
\vspace{-7mm}
\begin{equation} \label{49}
\end{equation}
\vspace{-7mm}
\[
\overline{\phi}_\tau(x)=\sum_{p}\left[a^+_{\tau,p}\overline{\phi^{(+)}_{\tau}}(x)
+ b_{\tau,p}\overline{\phi^{(-)}_{\tau}}(x)\right]
\]
where positive and negative parts of the wave function are given by
Eq. (37). The creation and annihilation operators of particles
$a^+_{\tau,p}$, $a_{\tau,p}$, and creation and annihilation
operators of antiparticles $b^+_{\tau,p}$, $b_{\tau,p}$ obey the
commutation relations:
\[
[a_{\tau,p},a^+_{\tau',p'}]=\delta_{\tau\tau'} \delta_{pp'}
,~~~[a_{\tau,p},a_{\tau',p'}]=[a^+_{\tau,p},a^+_{\tau',p'}]=0 ,
\]
\begin{equation}
[b_{\tau,p},b^+_{\tau',p'}]=\delta_{\tau\tau'} \delta_{pp'}
,~~~[b_{\tau,p},b_{\tau',p'}]=[b^+_{\tau,p},b^+_{\tau',p'}]=0 ,
\label{50}
\end{equation}
\[
[a_{\tau,p},b_{\tau',p'}]=[a_{\tau,p},b^+_{\tau',p'}]
=[a^+_{\tau,p},b_{\tau',p'}]=[a^+_{\tau,p},b^+_{\tau',p'}]=0 .
\]
With the aid of Eq. (48)-(50), and normalization condition (39),
we obtain the Hamiltonian
\begin{equation}
H=\int {\cal H}d^3 x=\sum_{\tau,p}p_0\left(a^+_{\tau,p}
a_{\tau,p}+b_{\tau,p} b^+_{\tau,p}\right) . \label{51}
\end{equation}
It is not difficult to find from Eq. (49)-(50) commutation
relations as follows:
\begin{equation}
[\phi_{\tau M(x)},\phi_{\tau N}(x')]=[\overline{\phi}_{\tau M}(x),
\overline{\phi}_{\tau N}(x')] =0, \label{52}
\end{equation}
\begin{equation}
[\phi_{\tau M}(x),\overline{\phi}_{\tau N}(x')]=N_{\tau MN}(x,x'),
\label{53}
\end{equation}
\[
N_{\tau MN}(x,x')=N^+_{\tau MN}(x,x')-N^-_{\tau MN}(x,x') ,
\]
\begin{equation}
N^+_{\tau MN}(x,x')=\sum_{p}\left(\phi^{(+)}_{\tau}(x)\right)_M
\left(\overline{\phi^{(+)}_{\tau}}(x')\right)_N  ,\label{54}
\end{equation}
\[
N^-_{\tau MN}(x,x')=\sum_{p}\left(\phi^{(-)}_{\tau}(x)\right)_M
\left(\overline{\phi^{(-)}_{\tau,}}(x')\right)_N .
\]
One obtains from Eq. (49):
\begin{equation}
N^\pm_{\tau MN}(x,x')=\sum_{p}\frac{m_\tau^2}{p_0
Vm}\left(v_{\tau}(\pm p)\right)_M\left(\overline{v}_{\tau}(\pm
p)\right)_N\exp [\pm ip(x-x')] ,
 \label{55}
\end{equation}
Taking into consideration Eq. (42), (45), and the relation
$p^2=-m_\tau^2$, we find from Eq. (55):
\[
N^\pm_{\tau MN}(x,x')=\sum_{p}\left\{
\frac{i(m_1+m_2)^4m_\tau^2\widehat{p}\left(i\widehat{p}\mp
m\right)}{2p_0Vmm_1^2m_2^2 \left(m_\tau^4-m_1^2m_2^2\right)}\left[
\widehat{p}^2+
 \frac{m_\tau^4}{(m_1+m_2)^2}\right]\right\}_{MN}
\]
\begin{equation}
\times \exp [\pm ip(x-x')]= \biggl\{
\frac{(m_1+m_2)^4m_\tau^2\left(\pm\rho_\mu\partial_\mu\right)
\left(\pm\rho_\mu\partial_\mu  \mp m\right)}{mm_1^2m_2^2
\left(m_\tau^4-m_1^2m_2^2\right)}\label{56}
\end{equation}
\[
\times \left[
 \frac{m_\tau^4}{(m_1+m_2)^2}- \left(\rho_\mu\partial_\mu\right)^2\right]\biggr \}_{MN}
 \sum_{p}\frac{1}{2p_0V}\exp[\pm ip(x-x')] ,
\]
With the help of the singular functions \cite{Ahieser}
\[
\Delta_+(x)=\sum_{p}\frac{1}{2p_0V}\exp
(ipx),~~~~\Delta_-(x)=\sum_{p}\frac{1}{2p_0V}\exp (-ipx),
\]
\[
\Delta_0 (x)=i\left(\Delta_+(x)-\Delta_-(x)\right),
\]
we obtain from Eq. (54), (56)
\[
N_{\tau MN}(x,x')=-i\biggl\{
\frac{(m_1+m_2)^4m_\tau^2\left(\rho_\mu\partial_\mu\right)
\left(\rho_\mu\partial_\mu- m\right)}{mm_1^2m_2^2
\left(m_\tau^4-m_1^2m_2^2\right)}
\]
\vspace{-6mm}
\begin{equation} \label{57}
\end{equation}
\vspace{-6mm}
\[
\times\left[
 \frac{m_\tau^4}{(m_1+m_2)^2}- \left(\rho_\mu\partial_\mu\right)^2\right]\biggr
 \}_{MN}\Delta_0 (x-x') .
\]
For the points $x$ and $x'$, which are separated by the space-like
interval $(x-x')>0$, the commutator $[\phi_M
(x),\overline{\phi}_N(x')]$ equals zero due to the properties of
the function $\Delta_0 (x)$ \cite{Ahieser}. If $t=t'$, $[\phi_M
(\textbf{x},0),\overline{\phi}_N(\textbf{x}',0)]=N_{\tau
MN}(\textbf{x}-\textbf{x}',0)$, and the function $N_{\tau
MN}(\textbf{x}-\textbf{x}',0)$ can be obtained from Eq. (57) with
the help of equations
\begin{equation}
\partial_0^{2n}\Delta_0 (x)|_{t=0}=0 ,~~~~\partial_m^{n}\Delta_0
(x)|_{t=0}=0 ,~~~~\partial_0 \Delta_0 (x)|_{t=0}= \delta
(\textbf{x}) , \label{58}
\end{equation}
where $n=1,2,3,...$. One may verify, using Eq. (41), the validity
of the equation
\begin{equation}
\left(\rho_\mu\partial_\mu +m\right)N^\pm_{\tau}(x,x')=0 .
 \label{59}
\end{equation}
The propagator of scalar fields (the vacuum expectation of
chronological pairing of operators) can be defined in our
formalism as
\[
\langle T\phi_{\tau M}(x)\overline{\phi}_{\tau
N}(y)\rangle_0=N^c_{\tau MN}(x-y)
\]
\vspace{-6mm}
\begin{equation} \label{60}
\end{equation}
\vspace{-6mm}
\[
=\theta\left(x_0 -y_0\right)N^+_{\tau MN}(x-y)+\theta\left(y_0
-x_0\right)N^-_{\tau MN}(x-y) ,
\]
where $\theta(x)$ is the well known theta-function. We obtain from
Eq. (56):
\[
\langle T\phi_{\tau M}(x)\overline{\phi}_{\tau N}(y)\rangle_0
=\biggl\{
\frac{(m_1+m_2)^4m_\tau^2\left(\rho_\mu\partial_\mu\right)
\left(\rho_\mu\partial_\mu- m\right)}{mm_1^2m_2^2
\left(m_\tau^4-m_1^2m_2^2\right)}
\]
\vspace{-6mm}
\begin{equation} \label{61}
\end{equation}
\vspace{-6mm}
\[
\times\left[
 \frac{m_\tau^4}{(m_1+m_2)^2}- \left(\rho_\mu\partial_\mu\right)^2\right]\biggr
 \}_{MN}\Delta_c (x-y),
\]
and the function $\Delta_c (x-y)$ is given by
\begin{equation}
\Delta_c (x-y)=\theta\left(x_0
-y_0\right)\Delta_+(x-y)+\theta\left(y_0 -x_0\right)\Delta_-(x-y)
. \label{62}
\end{equation}
Propagators (61) are finite only for $m_1\neq m_2$. It is seen
from Eq. (61) that propagators have different signs for $\tau=1$
and $\tau=2$. This means that one of states of the scalar field is
the ghost \cite{Hawking}.

\section{Conclusion}

We have formulated the scalar field equation with higher
derivatives in the form of the 10-component first order
relativistic wave equation. This equation describes scalar
particles possessing two mass states. It should be noted that the
second order equation for scalar fields with one mass state is
formulated in the 5-component matrix form \cite{Duffin},
\cite{Kemmer}. The relativistically invariant bilinear form, and
the Lagrangian were obtained, and this allowed us to find the
canonical energy-momentum tensor and density of the
electromagnetic current by the standard procedure. We found the
quantum-mechanical Hamiltonian by the separation of dynamical and
non-dynamical components of the wave function. The wave function
entering the Hamiltonian equation possesses four components to
describe two mass states of scalar fields with positive and
negative energies. The first order relativistic wave equation as
well as the Hamiltonian equation are convenient for different
applications. The density matrix (matrix-dyad) found can be used
for calculations of electromagnetic possesses.  The first order
formalism allowed us to quantize HD scalar fields in a simple
manner. The HD scalar field theory considered can by applied for a
model of inflation of the universe, where one of states of the
scalar field is identified with Quintessence and another, the
ghost, with Phantom \cite{Zhang}.

\section{Appendix}

With the help of Eq. (4), we find useful products of matrices
entering the Hamiltonian (34):
\begin{equation}
\rho_4 \rho_m =\frac{m}{m_1+m_2}\left(\sigma\varepsilon
^{4,m}+\varepsilon ^{\widetilde{4},m}-\varepsilon
^{4,\widetilde{m}}\right) ,\label{63}
\end{equation}
\begin{equation}
\rho_4^3 \rho_m
=\left(\frac{m}{m_1+m_2}\right)^2\left[\left(\sigma^2-1\right)\varepsilon
^{4,m}+\sigma\varepsilon ^{\widetilde{4},m}-\sigma\varepsilon
^{4,\widetilde{m}}-\varepsilon
^{\widetilde{4},\widetilde{m}}\right] ,
 \label{64}
\end{equation}
\begin{equation}
\rho_4 \rho_m
\Pi\rho_n=\delta_{mn}\left(\frac{m}{m_1+m_2}\right)^2\left(\varepsilon
^{4,\widetilde{0}}-\varepsilon
^{\widetilde{4},0}-\sigma\varepsilon ^{4,0}\right) ,
 \label{65}
\end{equation}
\begin{equation}
\rho_4^3 \rho_m
\Pi\rho_n=\delta_{mn}\left(\frac{m}{m_1+m_2}\right)^3\left[\varepsilon
^{\widetilde{4},\widetilde{0}}+\sigma\varepsilon
^{4,\widetilde{0}}-\sigma\varepsilon
^{\widetilde{4},0}+\left(1-\sigma^2\right)\varepsilon
^{4,0}\right] ,
 \label{66}
\end{equation}
\begin{equation}
\rho_4 \Pi=0 .
 \label{67}
\end{equation}
Using the definition $\hat{p}=p_\mu \rho_\mu$, we obtain:
\[
\hat{p}^2=\frac{mp^2}{m_1+m_2}\left(\sigma\varepsilon
^{0,0}+\varepsilon ^{\widetilde{0} ,0}-\varepsilon
^{0,\widetilde{0}}\right)
\]
\vspace{-7mm}
\begin{equation} \label{68}
\end{equation}
\vspace{-7mm}
\[
-\frac{m}{m_1+m_2}p_\mu p_\nu \left(\varepsilon
^{\mu,\widetilde{\nu}}-\varepsilon ^{\widetilde{\mu}
,\nu}-\sigma\varepsilon ^{\mu,\nu}\right),
 \]
\[
\hat{p}^3=\frac{mp^2}{m_1+m_2}p_\mu\left[\sigma\left(\varepsilon
^{0, \widetilde{\mu}}-\varepsilon ^{\widetilde{0}
,\mu}\right)+\left(1-\sigma^2\right)\varepsilon
^{0,\mu}+\varepsilon ^{
\widetilde{0},\widetilde{\mu}}\right]
\]
\vspace{-7mm}
\begin{equation} \label{695}
\end{equation}
\vspace{-7mm}
\[
+\frac{m^2p^2}{(m_1+m_2)^2}p_\mu \left(\varepsilon
^{\mu,\widetilde{0}}- \varepsilon
^{\widetilde{\mu},0}-\sigma\varepsilon ^{\mu, 0}\right),
\]
\begin{equation}
\hat{p}^4=\frac{mp^2\sigma}{m_1+m_2}\hat{p}^2
-\frac{m^2p^2}{(m_1+m_2)^2}\left[p^2\left(\varepsilon
^{0,0}+\varepsilon ^{\widetilde{0},\widetilde{0}}\right)+p_\mu
p_\nu \left( \varepsilon^{\mu,\nu}+\varepsilon
^{\widetilde{\mu},\widetilde{\nu}}\right)\right],
 \label{70}
\end{equation}
\begin{equation}
\hat{p}^5=
\frac{\left(m_1^2+m_2^2\right)p^2}{\left(m_1+m_2\right)^2}
\hat{p}^3 - \frac{m^2 p^4 }{\left(m_1+m_2\right)^2} \hat{p},
\label{71}
\end{equation}

\end{document}